\documentclass{jfm}

\usepackage{graphicx}
\usepackage{natbib}
\usepackage{amsmath}

\ifCUPmtlplainloaded \else
  \checkfont{eurm10}
  \iffontfound
    \IfFileExists{upmath.sty}
      {\typeout{^^JFound AMS Euler Roman fonts on the system,
                   using the 'upmath' package.^^J}%
       \usepackage{upmath}}
      {\typeout{^^JFound AMS Euler Roman fonts on the system, but you
                   dont seem to have the}%
       \typeout{'upmath' package installed. JFM.cls can take advantage
                 of these fonts,^^Jif you use 'upmath' package.^^J}%
      }
  \else
  \fi
\fi

\ifCUPmtlplainloaded \else
  \checkfont{msam10}
  \iffontfound
    \IfFileExists{amssymb.sty}
      {\typeout{^^JFound AMS Symbol fonts on the system, using the
                'amssymb' package.^^J}%
       \usepackage{amssymb}%
       \let\le=\leqslant  
         
      }{}
  \fi
\fi

\ifCUPmtlplainloaded \else
  \IfFileExists{amsbsy.sty}
    {\typeout{^^JFound the 'amsbsy' package on the system, using it.^^J}%
     \usepackage{amsbsy}}
    {}
\fi

\newsavebox{\astrutbox}
\sbox{\astrutbox}{\rule[-5pt]{0pt}{20pt}}

\usepackage{color}

\graphicspath{ {/} }

\title[Effect of morphology on an inverted flag]{Effect of morphology on the large-amplitude flapping dynamics of an inverted flag in a uniform flow}

\author[B. Fan, C.~Huertas-Cerdeira, J. Coss\'e, J.~E. Sader, and M. Gharib]{Boyu~Fan$^{1}$, Cecilia~Huertas-Cerdeira$^1$, Julia~Coss\'{e}$^1$, John E. Sader$^{2,3}$ and Morteza~Gharib$^1$}
\affiliation{$^1$Division of Engineering and Applied Science, California Institute of Technology, Pasadena, CA 91125, USA\\[\affilskip]
$^2$ARC Centre of Excellence in Exciton Science, School of Mathematics and Statistics, The University of Melbourne, Victoria 3010, Australia\\[\affilskip] $^3$Department of Physics, California Institute of Technology, Pasadena, CA 91125, USA\\[\affilskip]
}

\begin{document}

\maketitle

\begin{abstract}
The stability of a cantilevered elastic sheet in a uniform flow has been studied extensively due to its importance in engineering and its prevalence in natural structures. Varying the flow speed can give rise to a range of dynamics including limit cycle behaviour and chaotic motion of the cantilevered sheet. Recently, the `inverted flag' configuration---a cantilevered elastic sheet aligned with the flow impinging on its free edge---has been observed to produce large-amplitude flapping over a finite band of flow speeds. This flapping phenomenon has been found to be a vortex-induced vibration, and only occurs at sufficiently large Reynolds numbers. In all cases studied, the inverted flag has been formed from a cantilevered sheet of rectangular morphology, i.e., the planform of its elastic sheet is a rectangle. Here, we investigate the effect of the inverted flag's morphology on its resulting stability and dynamics. We choose a trapezoidal planform which is explored using experiment and an analytical theory for the divergence instability of an inverted flag of arbitrary morphology. Strikingly, for this planform we observe that the flow speed range over which flapping occurs scales approximately with the flow speed at which the divergence instability occurs. This provides a means by which to predict and control flapping. In a biological setting, leaves in a wind can also align themselves in an inverted flag configuration. Motivated by this natural occurrence we also study the effect of adding an artificial `petiole' (a thin elastic stalk that connects the sheet to the clamp) on the inverted flag's dynamics. We find that the petiole serves to partially decouple fluid forces from elastic forces, for which an analytical theory is also derived, in addition to increasing the freedom by which the flapping dynamics can be tuned. These results highlight the intricacies of the flapping instability and account for some of the varied dynamics of leaves in nature.
\end{abstract}


\section{Introduction} 									\label{intro_section}

\begin{figure}
	\centering
	\centerline{\includegraphics[width=1\columnwidth]{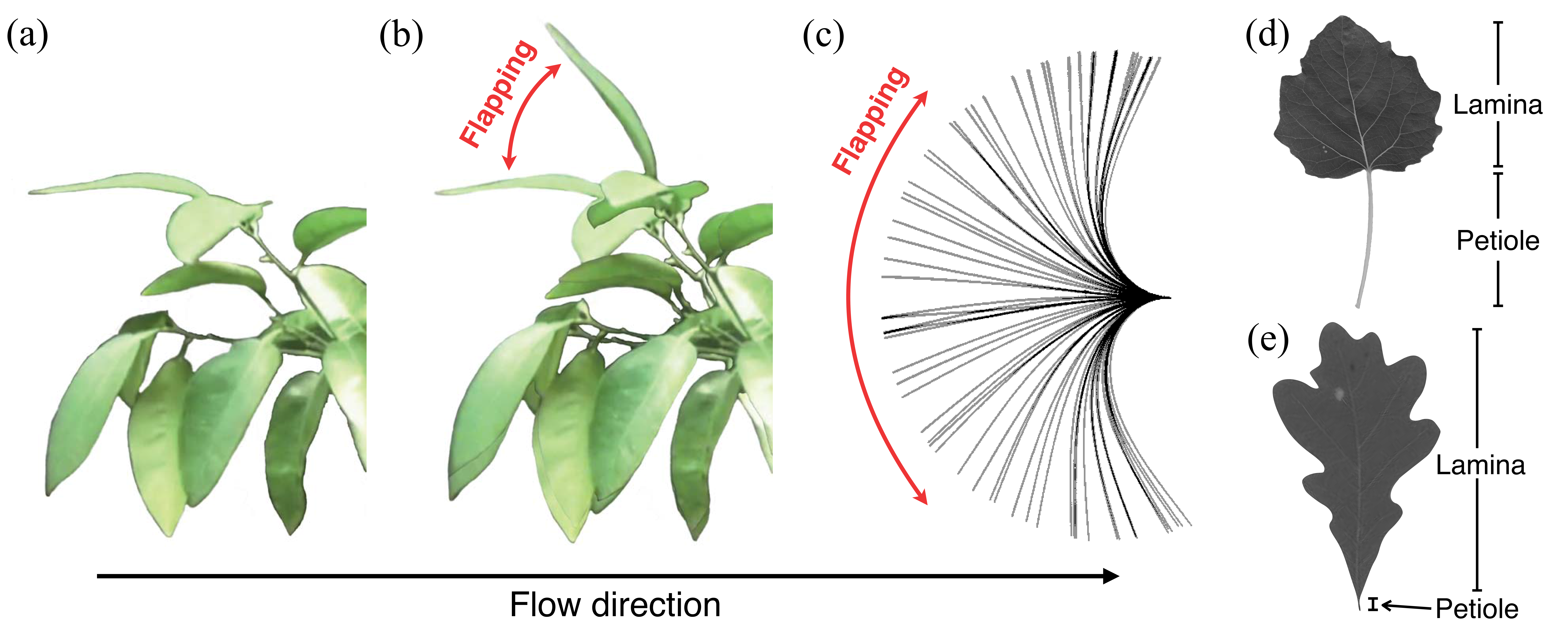}}
	\caption{(a) Leaves on a branch (image at an initial time), and (b) 0.67 seconds later with initial position overlaid to highlight flapping. (c) Stroboscopic image of an inverted flag exhibiting large-amplitude flapping instability from an edge-on view. Taken from present study: Flag E (standard rectangular inverted flag) at flow speed $U$ = 5.43 m/s; see Section~\ref{trapezoid}. (d) Typical white poplar leaf with lobed/heart-shaped base and long petiole, and (e) white oak leaf with acute base and short petiole (from Leafsnap dataset \cite{kumar12c}, not to scale).}
	\label{flap}
\end{figure}

As one of the simplest and most readily found examples of aeroelastic phenomena in nature, the dynamics of leaves excited by an impinging wind has contributed to the understanding of both plant biomechanics as well as fundamental aeroelasticity; see \cite{langre08a} for a review. The seminal work of \cite{vogel89a} introduced the now well-studied phenomena of drag reduction via reconfiguration of flexible bodies. However, most experimental and theoretical studies have focused on leaves oriented in the \textit{conventional} configuration of a flag fluttering in the wind, i.e., the flow impinges on the restrained end of the leaf with its free edge downstream. Even among leaves in a single tree, their location on the tree and thus orientation with respect to the impinging wind affect their observed dynamics. Thus far, no report has systematically studied the motion of leaves in an \textit{inverted} orientation, where the leading edge of the leaf is free and the trailing edge is fixed. While a na\"ive assessment would predict that the leaf would simply bend around the fixed end to return back to the conventional orientation, recent experimental work has shown that the dynamics are much more complex. Figures~\ref{flap}a and \ref{flap}b show a single measurement of such an inverted leaf (taken from \cite{sader16a}) where large-amplitude flapping (limit cycle behaviour) is evident. Detailed studies of the related and canonical `inverted flag' configuration---a cantilevered elastic sheet oriented such that the flow impinges on its free edge with its clamp downstream---provide an explanation for this leaf behaviour and show that orientation of the flag can have a dramatic effect~\citep{kim13a,cosse14i,tang15a,ryu15a,sader16a}.

\cite{kim13a} reported the first detailed measurements on the stability and dynamics of the inverted flag, in both air and water. The behaviour of the flag was characterised primarily by two dimensionless groups~\citep{kornecki76a,shelley11a}, a scaled flow speed:
\begin{equation}
\kappa \equiv \frac{\rho U^2 L^3}{D},
\label{kappa}
\end{equation}
and an added mass parameter:
\begin{equation}
\mu\equiv\frac{\rho L}{\rho_s h},
\label{muparam}
\end{equation}
where $\rho$ is the fluid density, $U$ the impinging flow speed, $L$ is the sheet length (in the direction of flow), the flexural rigidity of the elastic sheet $D\equiv E h^3/(12 [1-\nu^2])$, where $E$ and $\nu$ are its Young's modulus and Poisson's ratio, respectively, $h$ is its thickness, and $\rho_s$ is the sheet's density.  Note that the dimensionless flow speed, $\kappa$, represents a ratio of hydrodynamic forces to elastic restoring forces and can also be interpreted as a dimensionless reciprocal bending stiffness. The added mass parameter, $\mu$, specifies the relative importance of fluid-to-solid inertia.

A small range of aspect ratios, $H/L  = 1$ to 1.3, where $H$ is the sheet height (perpendicular to the flow) were considered by \cite{kim13a}. The inverted flag's dynamics were found to depend  predominantly  on the dimensionless flow speed, $\kappa$, with a variation in the added mass parameter, $\mu$, by a factor of 1000---from $\mu\approx 1$ (air) to 1000 (water)---producing only a moderate change in behaviour. The Reynolds number studied was in the range $\mathrm{Re}\equiv  U\! L / \nu \approx 10^4 - 10^5$, where $\nu$ is the fluid's kinematic viscosity.

It was observed that by increasing the flow speed from zero, the inverted flag's initial stationary state abruptly gave way to large-amplitude periodic flapping (see figure~\ref{flap}c) which eventually ceased (again abruptly) to a deflected stationary state. The value of $\kappa = \kappa_\mathrm{lower}$ at which large-amplitude flapping began (at low flow speed) was found to be identical in air and water. This was subsequently investigated by \cite{gurugubelli15a} using a computational method, which provided support for the proposal of \cite{kim13a} that large-amplitude flapping is initiated by a divergence instability. This was proved mathematically by \cite{sader16a} who conducted a linear stability analysis. \cite{sader16a} also reported results of an extensive investigation of the underlying physical mechanisms of the inverted flag's dynamics and concluded that large-amplitude flapping is a vortex-induced vibration (VIV). This finding was consistent with previous numerical simulations \cite{ryu15a} who observed that large-amplitude flapping does not occur for $\mathrm{Re} \lesssim 50$; which is strikingly similar to the numerical value for cessation of VIV by an elastic circular cylinder. This aspect has been investigated further by \cite{goza17}, again using direct numerical simulation, who found that flapping of the inverted flag persists for values of Re below $\mathrm{Re} \approx 50$, but only when the mass ratio is small, i.e., $\mu \ll 1$. This is strikingly similar to the behaviour of an elastic circular  cylinder, which is also discussed by \cite{goza17}.

The dimensionless flow speed, $\kappa = \kappa_\mathrm{upper}$, at which large-amplitude flapping ceases (at high flow speed) is found to depend on the added mass parameter, $\mu$, in contrast to the (lower) flow speed at which large-amplitude flapping is initiated. \cite{sader16a} noted that this observation is consistent with the existence of VIV where changing the added mass parameter produces a change in effective damping, and thus the flow speed at which desynchronisation occurs between vortex shedding and vibrational motion of the sheet. Increasing the added mass parameter, $\mu$, reduces the dimensionless flow speed, $\kappa$, where large-amplitude flapping ceases, albeit weakly.

A substantial body of work on the inverted flag has now been reported, involving experimental, computational and analytical methods. These include exploration of the inverted flag's energy harvesting potential \citep{shoele16a,orrego17a}, which seems particularly attractive given the large oscillation amplitudes that the inverted flag exhibits relative to a conventional flag. The interaction of two inverted flags has been reported and displays an interesting combination of synchronous and asynchronous dynamics depending on the flow speed and relative position, separation and length of the flags~\citep{huertas-cerdeira18,huang2018coupling,ryu2018flapping}. The aspect ratio, $H/L$,  of the inverted flag's elastic sheet has been found to play a critical role and strongly affect the (lower) flow speed at which large-amplitude flapping is initiated~\citep{cosse14i,sader16a}; the (higher) flow speed at which large-amplitude flapping ceases is relatively insensitive to aspect ratio. Indeed, reducing the aspect ratio to values less than unity results in a change in dynamics with elimination of the divergence instability (that initiates large-amplitude flapping) and replacement with a saddle node bifurcation~\citep{sader16a0,tavallaeinejad18}---where multiple linearly stable equilibrium states exist. Investigation of an inverted flag with a splitter plate attached to its trailing edge has also been reported~\citep{gurugubelli17}. In addition to direct numerical simulations at moderate Reynolds numbers, $\mathrm{Re} \sim O(100)$,  discussed above, computations using large eddy simulation (LES) at experimentally relevant Reynolds numbers, $\mathrm{Re}\sim O(10^5)$---which produce turbulent flows---have been performed~\citep{Gilmanov2015}.

One key aspect that is yet to be studied is the effect of the inverted flag's \textit{morphology}, i.e., the plan view shape of the flag's elastic sheet, on its resulting stability and dynamics. Such knowledge is particularly relevant to understanding the behaviour of leaves that naturally exhibit a wide array of shapes; see figures~\ref{flap}d and \ref{flap}e. All reports to date have used the same flag morphology of \cite{kim13a}: an elastic sheet of \textit{rectangular} planform. Here, we explore the effect of changing this shape by varying the geometric parameters of a \textit{trapezoidal} morphology for the inverted flag; see figure~\ref{setup}a. This increases the phase space of the resulting dynamics by adding a single shape parameter: the ratio of the free edge height, $\Delta$, to that at its clamp, $H$; for a rectangular sheet, we have $\Delta/H=1$. This trapezoidal inverted flag is studied experimentally in the wind tunnel used in \cite{kim13a} and by extending the divergence instability theory reported by \cite{sader16a} to an arbitrary planform that enables analytical solution. For the latter, we introduce a `generalised aspect ratio'~\citep{Jones1990,anderson06b}:
\begin{equation}
AR \equiv \frac{b^2}{S},
\label{AR}
\end{equation}
where $S$ is the surface area of the sheet's planform and $b$ is the maximum height of the sheet (perpendicular to the flow). 

Most leaves exhibit a short and narrow section---termed a `petiole'---that is attached to the branch of a tree (or to the tree itself); see figure~\ref{flap}d. The effect of this morphological feature on the dynamics of leaves with an impinging flow is also yet to be studied. To glean insight into its influence on the dynamics of leaves we  experimentally investigate inverted flags that contain such a petiole. This again increases the phase space of the resulting system. Therefore, in this initial study we focus on inverted flags (containing a petiole) with elastic sheets of rectangular planform only.

We begin by summarising details of the experimental apparatus and the measurement procedure in Section 2. This is followed in Section~\ref{trapezoid} by measurement data on  inverted flags of trapezoidal morphology that do not contain a petiole.  A detailed exploration of the underlying physical mechanisms and a discussion of the measurement data are provided. A general theory for the divergence instability of an inverted flag of arbitrary planform is presented in Section~\ref{theorysec} which includes comparison to the measurements reported in Section~\ref{trapezoid}. Finally, the effect of introducing a petiole to an inverted flag of rectangular planform is studied in Section~\ref{rectangle} together with commensurate theory.


\section{Experimental details}\label{expdet}

\begin{figure}
	\centering
	\centerline{\includegraphics[width=0.8\columnwidth]{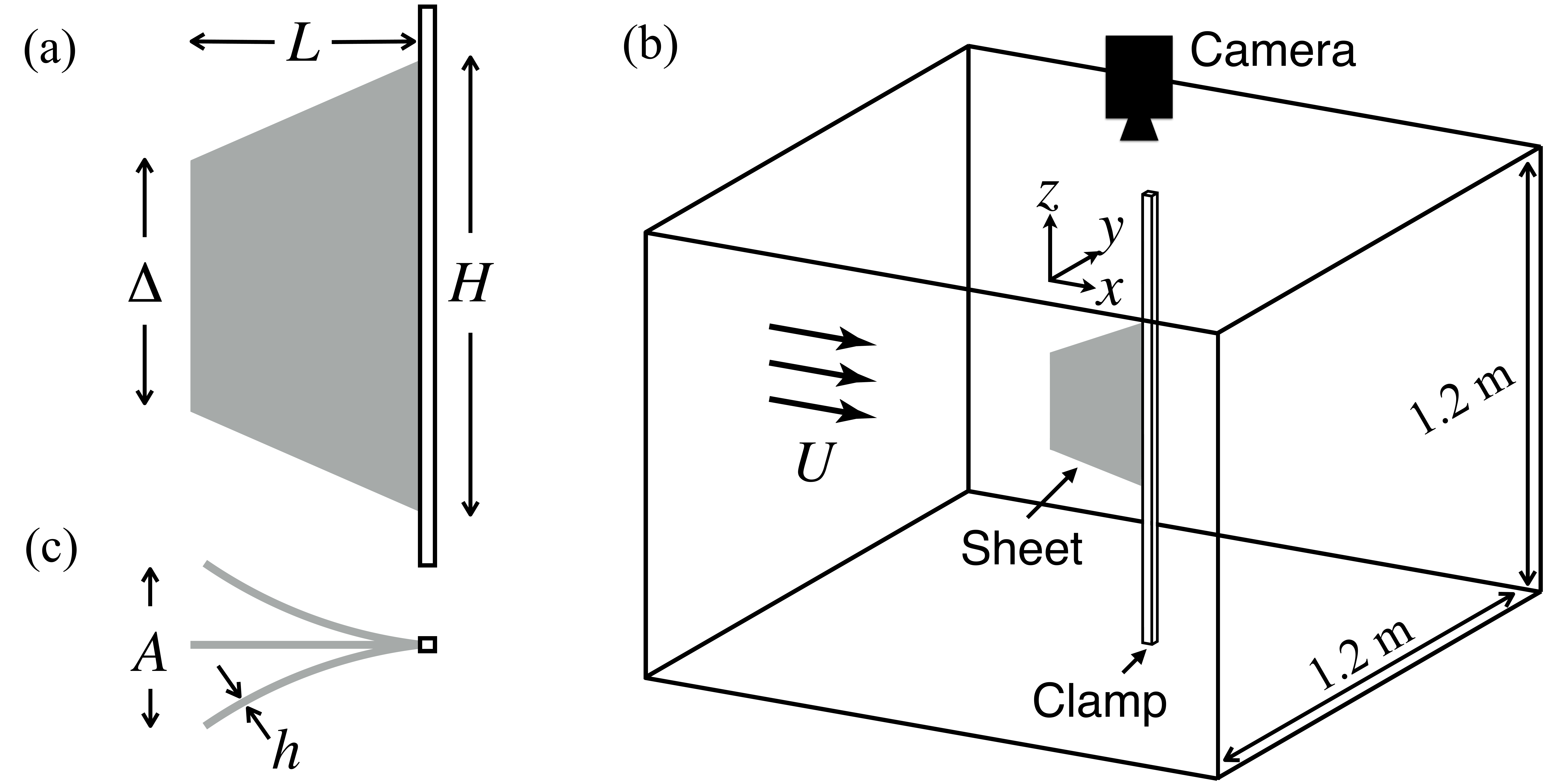}}
	\caption{(a) Schematic of the trapezoidal inverted flag (without a petiole) studied in Section \ref{trapezoid} from a side view perspective. (b) Experimental setup in wind tunnel showing Cartesian coordinate system whose origin is located at the middle of the sheet's leading free edge. (c) Top/camera view showing the straight sheet during flapping.}
	\label{setup}
\end{figure}

Inverted flags of trapezoidal (and rectangular) planforms are constructed from polycarbonate sheets with a Young's modulus $E=2.41$ GPa, Poisson's ratio $\nu=0.38$ and density $\rho_s = 1.2\times 10^3~\mathrm{kg/m^3}$. These flags are studied in an open loop wind tunnel with a square cross section of side length 1.2 m, which is capable of producing a free-stream flow speed, $U$, between 1.8 and 8.5 m/s; see figure~\ref{setup}b for a schematic of the wind tunnel. Further details on the wind tunnel are reported in \cite{kim13a,sader16a,huertas-cerdeira18}.

The inverted flag is oriented so that the impinging free stream is parallel to the flag's undeformed polycarbonate sheet. Data is collected by increasing the free stream speed from rest, and monitoring the inverted flag's dynamics at discrete speeds that are held constant for the measurement. A high-speed camera mounted above the test section captures the sheet motion at 100 -- 600 frames per second for varying flow speeds. Kinematic data of the top edge of the sheet are then extracted. See figure~\ref{setup}c for a schematic illustrating the camera viewpoint which shows the sheet thickness, $h$, and (peak-to-peak) oscillation amplitude, $A$, of a flapping sheet. We minimise the effects of gravity by clamping the sheet vertically. All deformations reported in this study, unless otherwise noted, are observed to be primarily two-dimensional, i.e., occurring the $xy$-plane; see figure~\ref{setup}b for the coordinate system. The natural resonant frequency of an inverted flag, $f_\mathrm{res} $, is measured by striking the sheet in still air and allowing it to return to its rest position.

\subsection{No petioles}\label{nopetioles}

We first consider inverted flags \textit{without} petioles, i.e., the trapezoidal sheet is attached directly to the clamp, formed from a single sheet of polycarbonate. The free stream impinges on the leading free edge of the sheet while the trailing edge of the sheet is clamped between two thin but rigid steel strips. The heights of all trapezoidal sheets are fixed at $H=19.05$ cm and their lengths are $L=15.24$ cm (giving $H/L=1.25$), which results in a Reynolds number of $\mathrm{Re}\sim O(10^4)$. The polycarbonate sheets have thicknesses of $h=0.25$ and 0.38 mm, corresponding to added mass parameters of $\mu=0.61$ and 0.40, respectively. The trapezoidal shape of the sheets are varied by adjusting the shape parameter, $\Delta/H$, between $\Delta/H=0$ and 1.75; the former corresponds to a triangular morphology.

\begin{figure}
	\centering
	\centerline{\includegraphics[width=0.8\columnwidth]{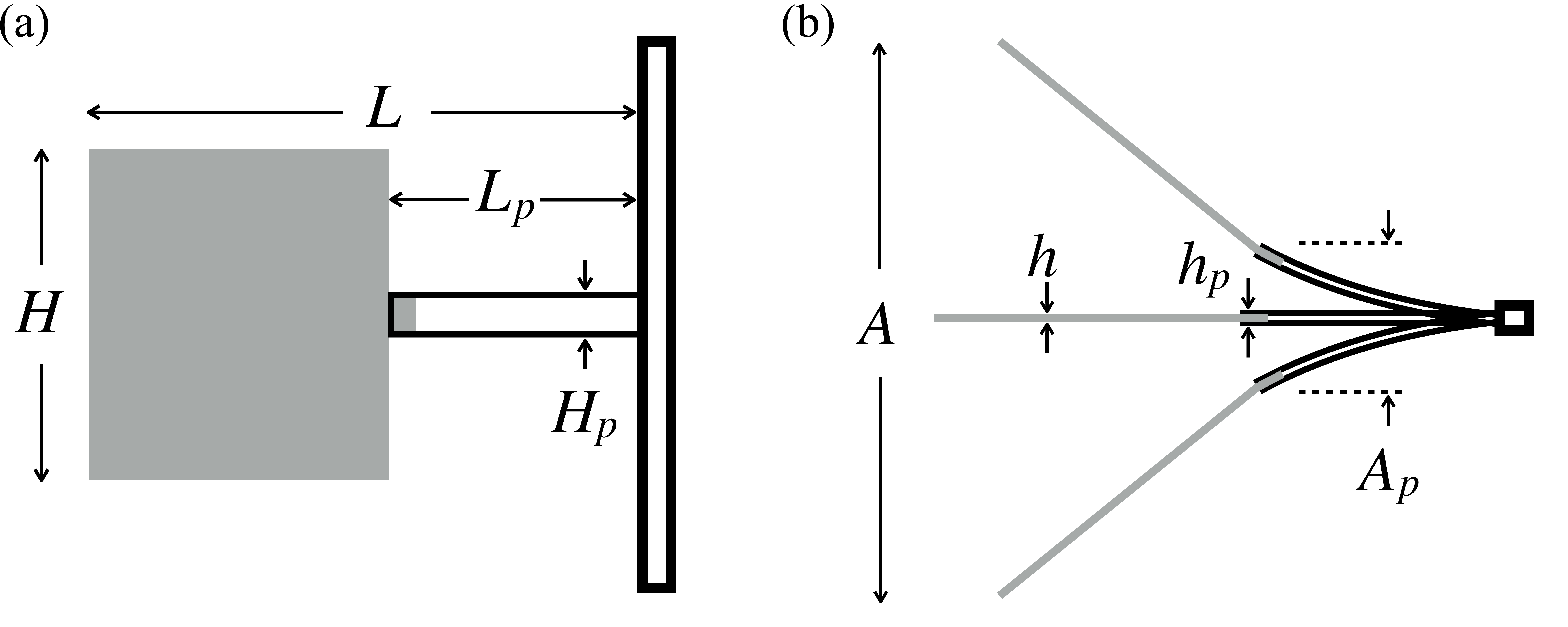}}
	\caption{Schematic of an inverted flag of rectangular morphology with a petiole. (a) Side view, and (b) top/camera view with straight rest position and flapping dynamics overlaid. The shaded area corresponds to the elastic polycarbonate sheet with a small tab-like overlap region used to join it to the petiole.}
	\label{petiole}
\end{figure}

\subsection{Petioles}\label{petioles}

The effect of including a petiole is studied for sheets of rectangular planform only. The elastic sheet and petiole are both fabricated from polycarbonate sheets. Each petiole is constructed by sandwiching two slender polycarbonate strips and bonding them with cyanoacrylate adhesive. The inverted flag is constructed from a single large rectangular sheet that is joined to the petiole using a small rectangular tab-like overlap region on the sheet; see figure~\ref{petiole}a. The resulting inverted flag is mounted in the wind tunnel by rigidly clamping the downstream end of the petiole to the steel strips (which are described in Section~\ref{nopetioles}) with the flow impinging on the sheet's free edge. The petiole length $L_{p}$, height $H_{p}$, and thickness $h_{p}$ are illustrated in figures~\ref{petiole}a and \ref{petiole}b. The petiole length, $L_p$, is varied relative to sheet length, $(L-L_p)$, while fixing total length of the inverted flag, $L$, height $H$, and the sheet thickness $h$. The petiole's flexural rigidity is adjusted by changing its thickness, $h_p$. Parameter values used in this study are given in Table~\ref{table}.

Also shown in Table~\ref{table} are measured values for the inverted flags' maximum flapping frequencies (driven by the impinging flow) and their natural resonant frequencies (in the absence of flow). The flapping frequency is determined via the average number of cycles per unit time, because the dynamics are highly periodic, and $f_\mathrm{max}$ is the maximum frequency of the flapping dynamics as the flow speed is varied  (excluding chaotic flapping).

\section{Trapezoidal inverted flag without a petiole}
\label{trapezoid}

\begin{figure}
	\centering
	\centerline{\includegraphics[width=1\columnwidth]{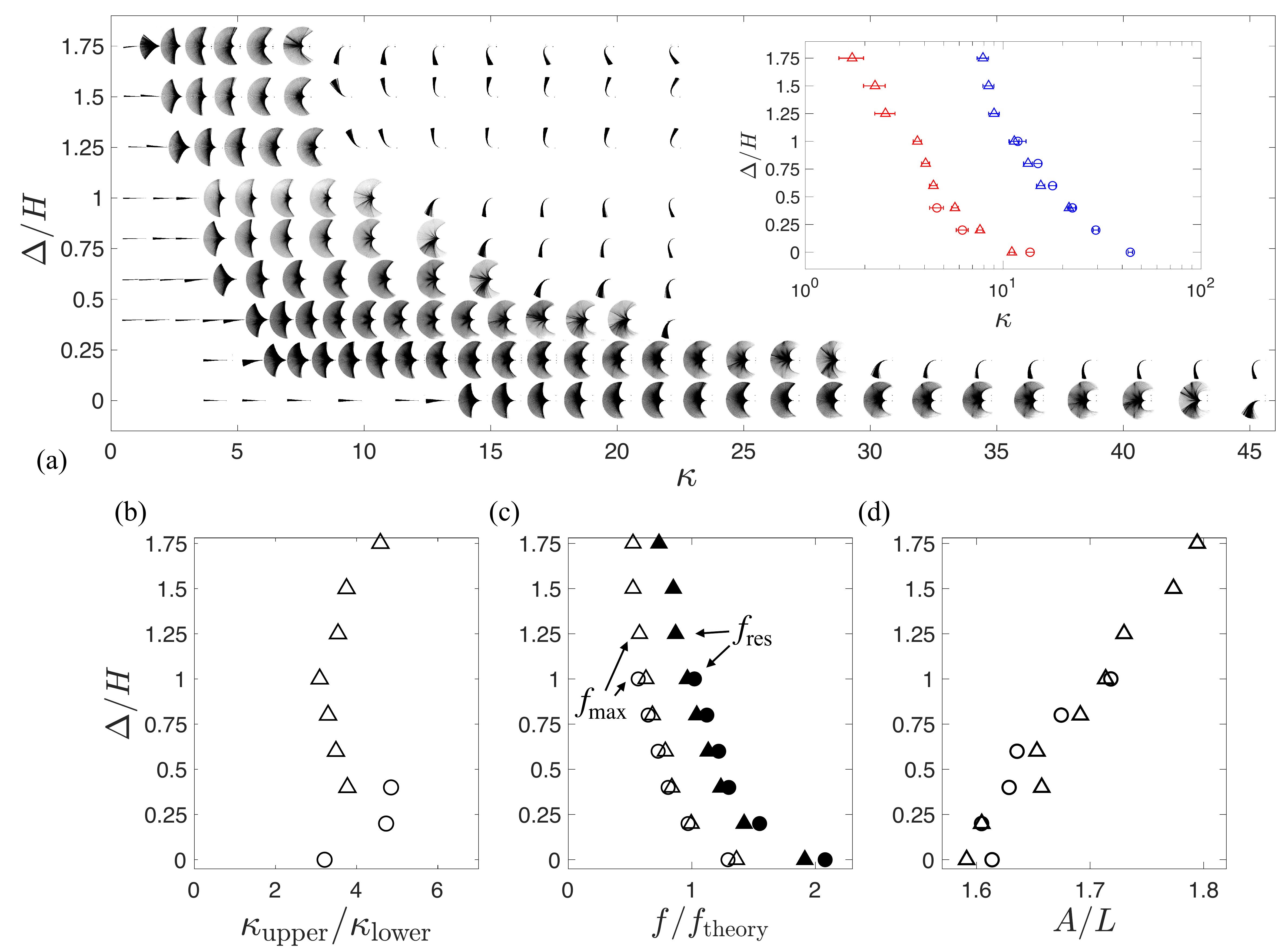}}
	\caption{Trapezoidal inverted flag behaviour as a function of shape parameter, $\Delta/H$, and the dimensionless flow speed, $\kappa$. (a) Stroboscopic images of the sheet motion are taken over 2.5 seconds and subsequently overlaid. Sheet dynamics with an added mass parameter of $\mu=0.61$ ($\circ$) are shown for $\Delta/H=0 $ and 0.2, while dynamics at $\mu=0.40$ ($\vartriangle$) are given for $\Delta/H>0.2 $; this enables all three behavioural regimes to be reported here. Inset gives the dimensionless flow speeds, $\kappa$, for the onset (left symbols) and cessation (right symbols) of large-amplitude flapping. (b) Plot of $\kappa_\mathrm{upper} / \kappa_\mathrm{lower}$, taken from data in the inset of (a)---shows this ratio is approximately constant. (c) Flapping frequencies, $f_\mathrm{max}$, and natural resonant frequencies, $f_\mathrm{res}$ \textcolor{black}{($\mu=0.61$ ($\bullet$) and $\mu=0.40$ ($\blacktriangle$)).} (d) Maximum  amplitude of trapezoidal inverted flags in flow speed range where flapping occurs. Inverted flags with two added mass parameters: $\mu=0.40$ ($\vartriangle$) and $\mu=0.61$ ($\circ$).}
	\label{lamina_results}
\end{figure}

In this section, we report measurements of the stability and dynamics of inverted flags with trapezoidal planforms that do not have petioles.
For each elastic sheet morphology, we observe the presence of three major behavioural regimes in order of increasing flow speed: a stable zero deflection equilibrium referred to as the `straight mode', a large-amplitude `flapping mode', and a stable deflected equilibrium referred to as the \textcolor{black}{`deflected mode';} see stroboscopic images in figure~\ref{lamina_results}a. These three regimes are separated by abrupt transitions at two critical non-dimensional flow speeds: (i) $\kappa_\mathrm{lower}$, where the sheet's zero deflection equilibrium undergoes a divergence instability, giving way to the flapping mode, and (ii) $\kappa_\mathrm{upper}$, where flapping ceases. The triangular sheet ($\Delta/H=0$) requires the highest non-dimensional flow speed to begin and end flapping.  As the shape parameter $\Delta/H$ is increased, these critical flow speeds decrease along with the flow speed range over which flapping occurs. We note that $\kappa_\mathrm{lower}$ is insensitive to the added mass parameter, $\mu$, even for sheets of non-rectangular geometry. This is consistent with $\kappa_\mathrm{lower}$ being due to a divergence instability, which is a static fluid-structure phenomenon.
\begin{table}
	\centering
	\begin{tabular}{c c c c c c c c c c c c}
		Flag & $L$ & $H$ & $h$ & $L_p$ & $H_p$ & $h_p$ & Petiole & $f_\mathrm{max}$ & $f_\mathrm{res}$ \\
		\hline\noalign{\smallskip}
		 A & 17.8 & 17.8 & 0.051 & 6.4  & 1.3 & 0.102 & short, high rigidity & 2.1 & 3.3 \\
		 B & 17.8& 17.8 & 0.051 & 6.4  & 1.3 & 0.076 & short, low rigidity & 1.5 & 2.6 \\
		 C & 17.8 & 17.8 & 0.051 & 11.4 & 1.3 & 0.102 & long, high rigidity & 2.3 & 4.0 \\
		 D & 17.8  & 17.8 & 0.051 & 11.4 & 1.3 & 0.076 & long, low rigidity & 1.5 & 2.7 \\
		 E & 17.8 & 17.8 & 0.051 & -   & -   & -  & no petiole & 2.4 & 3.5 \\
	\end{tabular}
	\caption{Dimensions, petiole description and measured frequencies of rectangular inverted flags with a petiole; Flag E is a rectangular inverted flag with no petiole that is used for reference. These flags are studied in Section \ref{rectangle}. Here, $f_\mathrm{max}$ is the measured flapping frequency with an impinging flow and $f_\mathrm{res}$ is its measured natural frequency in the absence of flow. All length/height/thickness values are reported in cm and frequencies are  in Hz.}
	\label{table}
\end{table}

Strikingly, we observe that $\kappa_\mathrm{upper}/\kappa_\mathrm{lower}\approx 4$ for all trapezoidal morphologies; see figure~\ref{lamina_results}a (inset) and figure~\ref{lamina_results}b. This observation shows that the flow speed range over which flapping occurs can be predicted solely from knowledge of $\kappa_\mathrm{lower}$ for trapezoidal flags---which can in turn be theoretically calculated; see Section~\ref{theorysec}.

The flapping mode can be subdivided into (i) a (limit cycle) steady flapping mode, and (ii) a chaotic flapping mode that arises only at flow speeds close to and just below the flapping-deflected mode transition, $\kappa_\mathrm{upper}$. A steady flapping mode is evident as a smooth gradient in the stroboscopic images of figure~\ref{lamina_results}a, while aperiodic (chaotic) flapping displays prominent sharp edges. We observe that the range of $\kappa$ for chaotic flapping is widest for the triangular lamina ($\Delta/H=0$) and becomes narrower as the shape parameter $\Delta/H$ increases. For the rectangular sheets ($\Delta/H=1$), our observations are consistent with previous experimental observations made by \cite{kim13a} and \cite{sader16a}, as well as with the theoretical prediction of $\kappa_\mathrm{lower}\approx4.14$ using (2.15) of \cite{sader16a0}; which is valid for rectangular planforms of arbitrary aspect ratio.

To investigate the effect of the elastic sheet's morphology on the measured (i) flapping frequency, $f_\mathrm{max}$, and (ii)  natural resonant frequency, $f_\mathrm{res}$, and to compare this effect across sheets with different added mass parameters, $\mu$, we scale all frequencies by the theoretical natural resonant frequency  for a cantilevered rectangular sheet in an inviscid fluid~\citep{shen16a},
\begin{equation}\label{inviscid}
f_\mathrm{theory}=\frac{C_1^2}{2\pi L^2}\sqrt{\frac{D}{\rho_s h(1+a\mu)}},
\end{equation} 
where $C_1=1.875$ (for the fundamental flexural mode), $(1+a\mu)$ is the added apparent mass term that accounts for the fluid's inertial load, and $a$ is an $O(1)$ dimensionless coefficient that depends on the sheet's aspect ratio, $H/L$, the geometry of its clamp, and to a lesser degree the added mass parameter, $\mu$. Modelling the rigid steel strips that hold the flag's elastic sheet as a `line clamp',~\cite{shen16a} gives $a=0.33$ for all added mass parameters studied.  Equation~(\ref{inviscid}) shows that the air surrounding the sheet reduces its natural resonant frequency by between 6 and 10\% for the added mass parameters studied. It provides a natural frequency scale for the inverted flag.

Figure~\ref{lamina_results}c shows that (\ref{inviscid}) \textcolor{black}{captures the dominant behaviour of} the natural frequencies, $f_\mathrm{res}$, of all trapezoidal sheets studied (filled shapes), \textcolor{black}{as their shape is modified. This} is not  surprising because changing the trapezoidal shape (via $\Delta/H$) alters the mass and stiffness of the sheet in a commensurate manner. As the shape parameter, $\Delta/H$, of the trapezoidal sheet increases, both the flapping and natural resonant frequencies decrease monotonically with the flapping frequency always being less than the natural frequency.
The latter observation is consistent with previous measurements on an inverted flag of rectangular morphology~\citep{sader16a}; the flapping mode is associated with strong vortex shedding which can significantly modify the fluid load.
Strikingly, all measured flapping frequencies scaled by (\ref{inviscid}) are $O(1)$ quantities, 
\textcolor{black}{as expected for a vortex-induced vibration under moderate fluid loading, $\mu < 1$; see Section~\ref{expdet}~\citep{williamson04a,sader16a}.}

All trapezoidal inverted flags begin flapping at a Strouhal number, $\text{St} \equiv fA/U$ (following \cite{kim13a}), between 0.15 and 0.22, \textcolor{black}{that decreases with increasing flow speed to a value in the vicinity of 0.05--0.08 before the deflected mode emerges. This is} consistent with general observations of vortex-induced vibrations; as noted by \cite{sader16a} for inverted flags of rectangular planform. Strikingly, the maximum non-dimensional flapping amplitude, $A/L$, in the flapping regime increased approximately linearly with $\Delta/H$; see figure~\ref{lamina_results}d. \textcolor{black}{This is due to the expected slight difference in the vibrating mode shape as a function of flag morphology, which alters the maximum vertical extension (amplitude) during flapping; the difference is approximately 10\%.}

These measurements have implications to the dynamics of leaves in nature. Based solely on the sheet's morphology and with all else being held constant, our measurements suggest that white poplar leaves (see figure~\ref{flap}d) in an inverted configuration---that are geometrically analogous to $ \Delta/H \gg 1 $---would (i) undergo flapping at relatively high flow speeds, and (ii) for a wide range of flow speeds, with (iii) higher flapping frequency and lower amplitude. White oak leaves (see figure~\ref{flap}e)---analogous to $ \Delta/H \ll 1 $---on the other hand, would either undergo large-amplitude flapping at relatively low flow speeds and only for a narrow range of speeds, or simply transition directly from the straight mode to the deflected mode and not exhibit any flapping dynamics~\citep{sader16a0}.

\section{Theory for inverted flags of arbitrary morphology}\label{theorysec}

\begin{figure}
	\centering
	\centerline{\includegraphics[width=0.5\columnwidth]{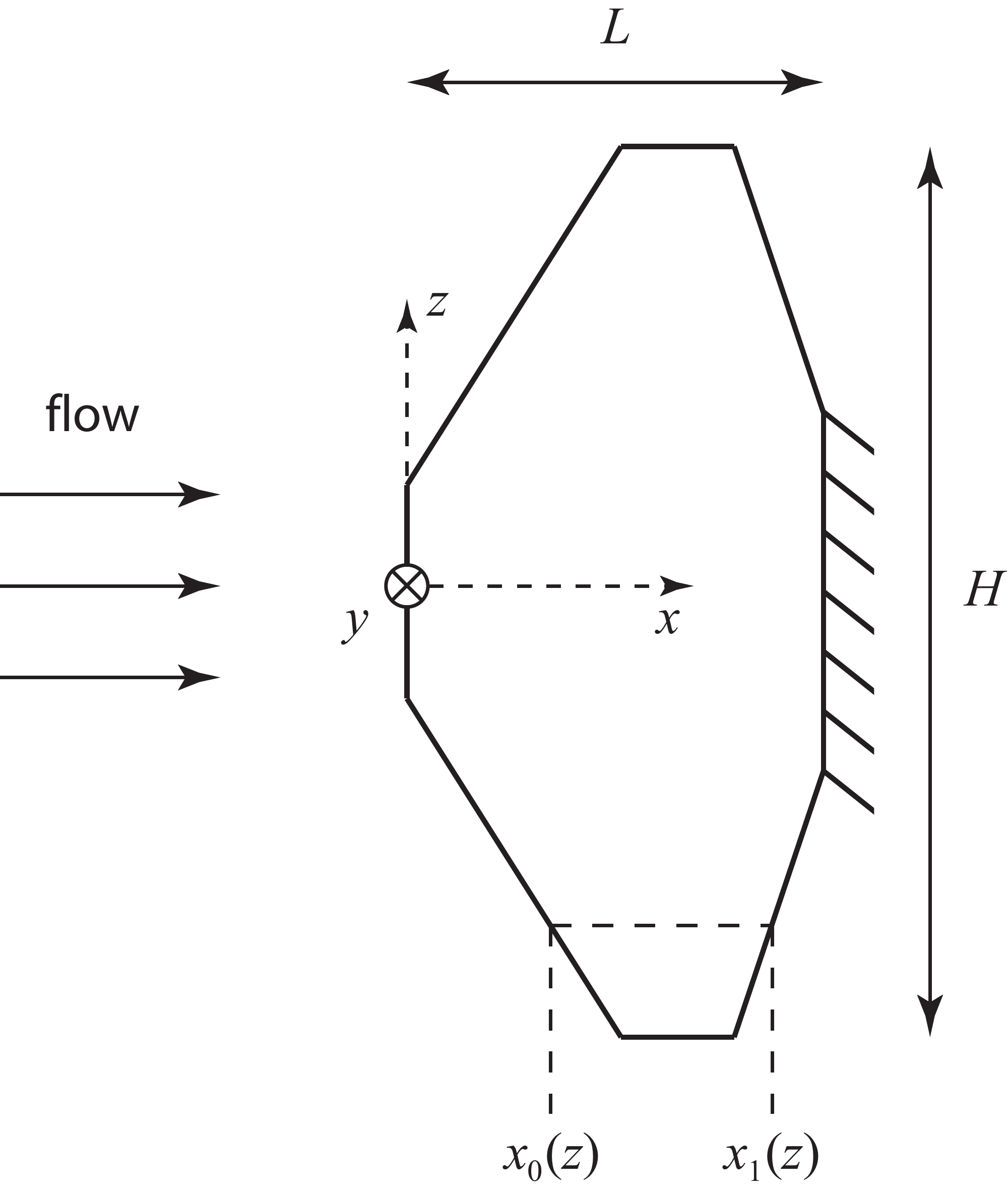}}
	\caption{Schematic  of an inverted flag of arbitrary morphology. Coordinate system, its origin (centre of leading edge), and the geometric parameters used in general theoretical model for a divergence instability in Section~\ref{theorysec} are shown.}
	\label{arbitraryflag}
\end{figure}

In this section, we present a general theoretical model for the divergence instability (that initiates large-amplitude flapping) of an inverted flag of \textit{arbitrary} morphology; see figure~\ref{arbitraryflag}. It is assumed that the flag's elastic sheet is of constant thickness and has uniform material properties. This generalises the theoretical framework of \cite{sader16a}, which was developed for inverted flags of rectangular planform, and simplifies it by enabling analytical calculation. This general theory is applied to inverted flags of trapezoidal planform which is then  compared to the measurements reported in Section~\ref{trapezoid}.

The general theory for arbitrary planforms is first formulated in the asymptotic limit, $\mathrm{AR} \rightarrow \infty$; see (\ref{AR}), which is then corrected for finite generalised aspect ratio, $\mathrm{AR}$. This is achieved using a strip model that utilises the \textcolor{black}{(quasi-steady)} analytical solution of \cite{kornecki76a} for a two-dimensional (rectangular) cantilevered plate and Prandtl's lifting line theory~\citep{Jones1990,anderson06b}, respectively. As such, the derived theory implicitly ignores the nonlinear effects of vortex lift---driven by vortex shedding along the side edges of the sheet---which manifest themselves with increasing importance as the generalised aspect ratio AR is reduced; see \cite{sader16a0}. Such nonlinear lift mechanisms are expected to enhance the total lift, and thus reduce the flow speed at which the divergence instability occurs~\citep{sader16a0}. Therefore, the presented theory  provides an upper bound on the dimensionless flow speed, $\kappa_\mathrm{lower}$.

\subsection{Strip model for arbitrary planform $(\mathrm{AR} \rightarrow \infty)$}\label{strip}

In the limit, $\mathrm{AR} \rightarrow \infty$, as defined in (\ref{AR}), the pressure distribution along the cantilevered sheet can be rigorously calculated by dividing the sheet into infinitesimal rectangular strips in the direction of flow (at each spanwise location). This limit corresponds to a sheet that is infinitely wide relative to its length. The dimensionless pressure jump distribution (scaled by $\rho U^2 A_\mathrm{disp}/L$, where $A_\mathrm{disp}$  is the magnitude of the sheet's free end \textcolor{black}{static displacement}) across each strip is then given by that of an infinitely wide (2D) rectangular cantilevered sheet~\citep{kornecki76a},
\begin{equation}
p(x,z)=  \frac{g(\theta, z) - g(0, z)}{\sin \theta},
\label{kornecki}
\end{equation}
with
\begin{subequations}
\label{korneckidetail}
\begin{gather}
\displaystyle g(\theta,z)=-\frac{2}{\pi} \int_0^\pi \frac{\partial w}{\partial \zeta} \frac{\sin^2 \phi}{\cos \phi - \cos \theta} d \phi , \\
\displaystyle x = x_0(z) + \frac{x_1(z)-x_0(z)}{2} (1 + \cos \theta) , \\
\displaystyle \zeta = x_0(z) + \frac{x_1(z)-x_0(z)}{2} (1 + \cos \phi) ,
 \end{gather}
\end{subequations}
where $\theta, \phi \in [0, \pi]$, $w(x,z)$ is the sheet's dimensionless displacement (scaled by $A_\mathrm{disp}$) normal to its plane, $x$ and $y$ are the dimensionless Cartesian coordinates (scaled by the sheet's length, $L$) parallel and normal to the flow direction, respectively, $x_0(z)$ and $x_1(z)$ are the $x$-coordinates of the leading and trailing edges of the strip located at $z=z$, respectively; see figure~\ref{arbitraryflag}. Note that there is a unique one-to-one relationship between $x$ and $\theta$, and also $\zeta$ and $\phi$.

To determine the dimensionless flow speed, $\kappa_\mathrm{lower}$, at which divergence occurs we make use of the small-deflection theory for thin plates~\citep{Landau1970},
\begin{equation}
\nabla^4 w (x, z) = \kappa \, p (x,z).
\label{plateequation}
\end{equation}
Equation~(\ref{plateequation}) is then multiplied by $w(x,z)$ and integrated over the sheet's surface, $S$, while imposing the sheet's boundary conditions and noting that $w (x, z)$ varies slowly with respect to $z$ (when $H/L \gg 1$), to give
\begin{equation}
\kappa_\mathrm{lower}^{\infty} = \frac{\displaystyle \int_S \left(\displaystyle \frac{\partial^2 w}{\partial x^2}\right)^2 dS}{\displaystyle \int_S p\, w \, dS},
\label{kappavariational}
\end{equation}
where the superscript, $\infty$, denotes use of a strip model and is valid for $\mathrm{AR} \rightarrow \infty$. Since the pressure, $p$, is linearly dependent on the displacement, $w$, it then follows that (\ref{kappavariational}) is independent of the displacement magnitude, $A_\mathrm{disp}$, as required.

In \cite{sader16a}, the deflection function, $w$, is determined using a Rayleigh-Ritz approach where $w$ is expressed as a general power series in $x$; the unknown coefficients are evaluated from the stationary point of (\ref{kappavariational}). To simplify implementation, we make use of the known deflection function for a \textcolor{black}{rectangular cantilevered sheet ($\Delta/H=1$)} under a static point load at its free end,
\begin{equation}
w(x,z)=\frac{1}{2} \left(2+x\right)\left(1-x\right)^2,
\label{cantdef}
\end{equation}
which enforces the free end condition at $x=0$ and the clamp at $x=1$. This enables analytical solution of $\kappa_\mathrm{lower}^{\infty}$ for sheets of arbitrary planform by substituting (\ref{kornecki}) and (\ref{cantdef}) into (\ref{kappavariational}). Because (\ref{cantdef}) is an approximation to the true deflection function \textcolor{black}{of a sheet of arbitrary morphology}, this approach guarantees that the resulting value for $\kappa_\mathrm{lower}^{\infty}$ is strictly an upper bound~\citep{Leissa1969}. A general power series expansion can be used (as in \cite{sader16a}) at the expense of increased complexity. However, because the sheet's true mode shape will be qualitatively identical to (\ref{cantdef}), and (\ref{kappavariational}) defines the Rayleigh quotient for the problem, use of a general power series is expected to produce only a moderate improvement in accuracy and is not implemented.

Substituting (\ref{cantdef}) into (\ref{kornecki}) gives the required pressure distribution,
\begin{equation}
p(x,z)=-\frac{3 \sin \theta}{8\! \left(1+\cos \theta\right)} \left[8\! \left(x_1^2-1\right) + 2\! \left(x_1-x_0\right) \left(x_0+3 x_1\right) \cos \theta + \left(x_0-x_1\right)^2 \cos 2 \theta  \right] ,
\label{pressure}
\end{equation}
where $x_0$,  $x_1$, and $\theta$ are defined in (\ref{korneckidetail}).  Equations~(\ref{kappavariational}) -- (\ref{pressure}), together with the defined relationship between $x$ and $\theta$ in (\ref{korneckidetail}), enable  the divergence instability of an inverted cantilever sheet of arbitrary planform to be calculated directly and trivially. \textcolor{black}{Care must be exercised when the expected deflection mode shape deviates significantly from (\ref{cantdef}); an example is considered in Section~\ref{trapezoidformula}.}

\subsection{Arbitrary planforms of finite aspect ratio}\label{theoryAR}

To account for finite generalised aspect ratio, AR, the approach reported by \cite{sader16a} is now applied to the general 2D strip theory of the previous section. For a flat sheet of arbitrary generalised aspect ratio, AR, Prandtl's lifting line theory~\citep{Jones1990,anderson06b} predicts a reduction in its lift slope by factor of $\approx (1+2/\mathrm{AR})$, ignoring the usual (small) Glauert parameter. This reduction in lift is due to the generation of side edge vortices that are always present on wings of finite aspect ratio. Therefore, the dimensionless flow speed at which divergence occurs for an inverted flag of arbitrary shape, relative to its two-dimensional value, $\kappa_\mathrm{lower}^{\infty}$,  is enhanced by this factor, giving the required result,
 \begin{equation}\label{Prandtl}
\kappa_\mathrm{lower} \approx \kappa_\mathrm{lower}^\infty \left(1+\frac{2}{\mathrm{AR}}\right),
\end{equation}
where $\kappa_\mathrm{lower}^\infty$ for a trapezoidal morphology is specified in (\ref{trapinfty}) below. Note that for a generalised aspect ratio, $\mathrm{AR}\sim O(1)$, which is typical in the  measurements reported in Section~\ref{trapezoid}, the value of $\kappa_\mathrm{lower}$ is approximately three times larger than that for an infinitely wide (2D) inverted flag. This shows that side edge vortices play a important role in controlling the stability of inverted flags of finite aspect ratio.

\textcolor{black}{A comparison of the predictions of (\ref{Prandtl}) to those obtained using a vortex lattice method (VLM)~\citep{Tornado135,sader16a} are given in figure~\ref{trapinftyfig}a. This is performed for an aspect ratio of $H/L=1.25$ which is studied experimentally in Section~\ref{trapezoid}. VLM simulations involve calculating the lifting slope of a rigid  trapezoidal sheet as a function of its shape parameter, $\Delta/H$; they provide a more accurate and rigorous approach in place of (\ref{Prandtl}), if needed. The data in figure~\ref{trapinftyfig}a shows that (\ref{Prandtl}) provides a good approximation.}

\begin{figure}
	\centering
	\centerline{\includegraphics[width=0.6\columnwidth]{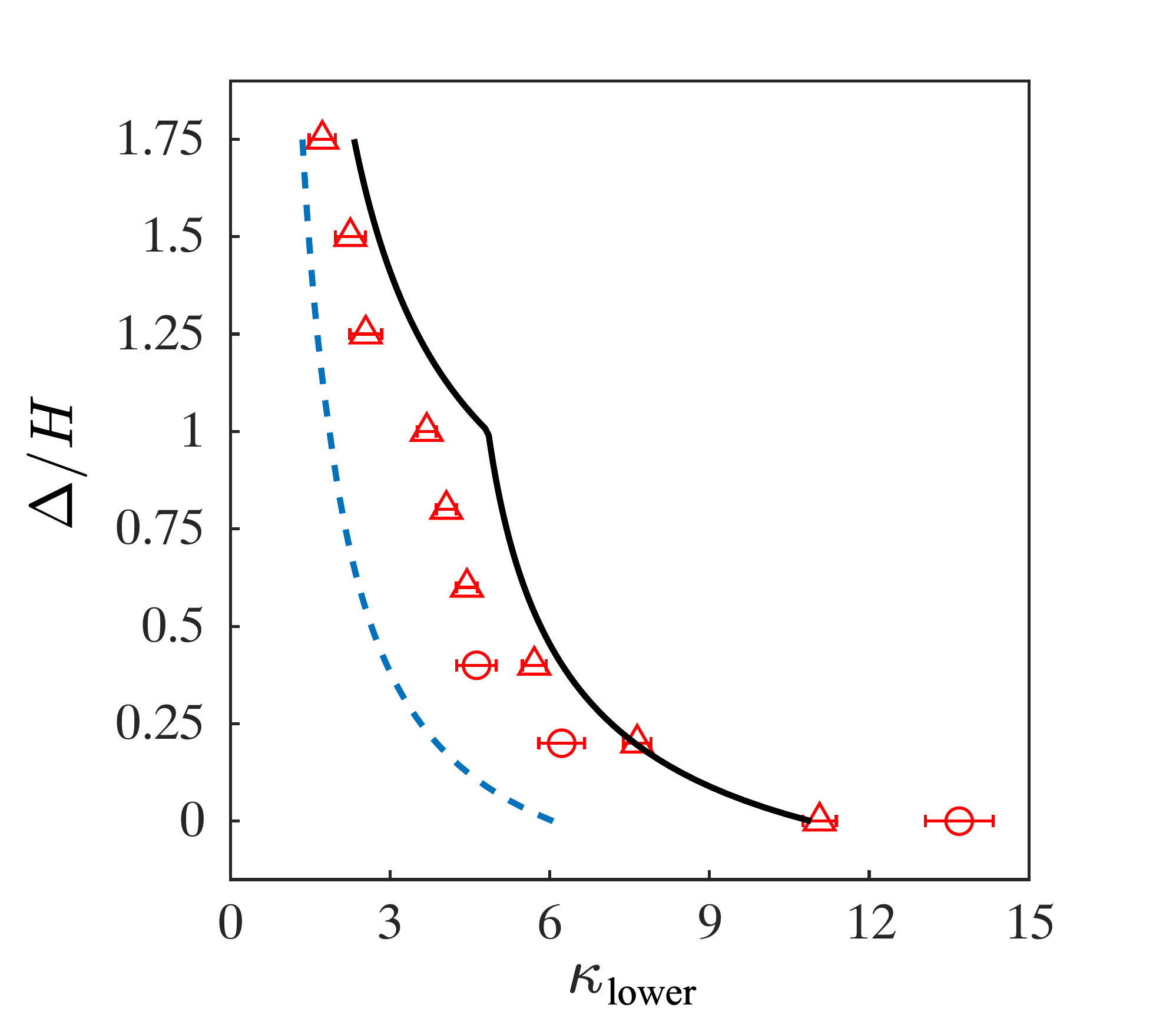}}
	\caption{Comparison of dimensionless flow speed at divergence, $\kappa_\mathrm{lower}$, obtained experimentally  for two added mass parameters: $\mu=0.40$ ($\vartriangle$) and $\mu=0.61$ ($\circ$) (see Section~\ref{trapezoid}), using the theory for infinite aspect ratio in (\ref{trapinfty}) (dashed line) and with the finite aspect ratio correction in (\ref{Prandtl}) (solid line). Error bars  quantify the uncertainty in measuring $\kappa_\mathrm{lower} $ for a single sheet, and define the 25\% and 75\% points discussed in Section~\ref{measuredetail}.}
	\label{theory}
\end{figure}

\textcolor{black}{We note that lifting line theory---which is used to derive (\ref{Prandtl})---makes no distinction between swept and symmetric wings, and is
normally applied to symmetric wings only~\citep{Jones1990,anderson06b}. Even so, (\ref{Prandtl}) captures the VLM simulations well. It is also interesting that the curves in figure~\ref{trapinftyfig}a present a discontinuous derivative at $\Delta/H=1$. This is consistent with the kink experimentally observed in the critical velocity at the same value of $\Delta/H$; see figures~\ref{lamina_results}a and \ref{theory}.}

\subsection{Trapezoidal planform}\label{trapezoidformula}

The general strip theory of Section~\ref{strip} is now used to calculate $\kappa_\mathrm{lower}^{\infty}$ for an inverted flag of trapezoidal morphology; the corresponding result for finite $\mathrm{AR}$ is trivially obtained using (\ref{Prandtl}). Calculation of $\kappa_\mathrm{lower}^{\infty}$ simply requires specification of the trapezoidal planform and integration of (\ref{kappavariational}), which yields
\begin{equation}
\kappa_\mathrm{lower}^{\infty} =
   \left\{
   \begin{aligned}
\frac{768}{\pi}\frac{3+\frac{\Delta}{H}}{121 + 404  \frac{\Delta}{H}}, \qquad & 0 \le \frac{\Delta}{H} < 1,\\\\
\frac{1792}{\pi}\frac{3+\frac{\Delta}{H}}{186+1039 \frac{\Delta}{H}}, \qquad& 1 \le \frac{\Delta}{H} \lesssim O(10),\\
   \end{aligned}
   \right.
   \label{trapinfty}
\end{equation}
\textcolor{black}{where origin of the $O(10)$ upper limit is explained below.} Equation~(\ref{trapinfty}) gives $\kappa_\mathrm{lower}^{\infty}=1.86$ for an inverted flag of rectangular morphology ($\Delta/H=1$), which is \textcolor{black}{slightly larger} than the exact value of $\kappa_\mathrm{lower}^{\infty}=1.85$~\citep{sader16a}, demonstrating the validity of the approximations implemented.

\begin{figure}
	\centering
	\centerline{\includegraphics[width=\columnwidth]{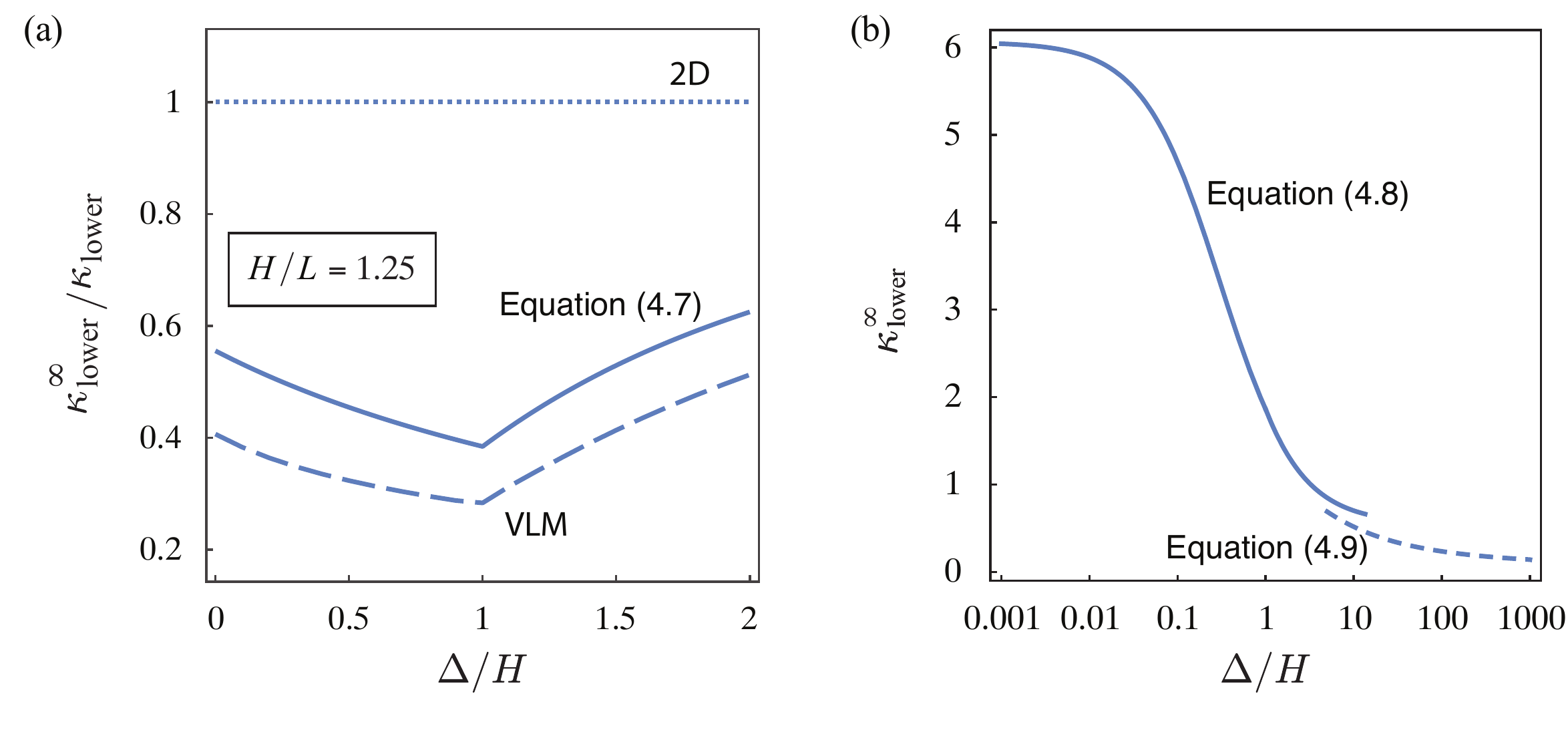}}
	\caption{\textcolor{black}{Dimensionless flow speed, $\kappa_\mathrm{lower}$, for the divergence instability of a trapezoidal inverted flag; superscript $\infty$ indicates  infinite aspect ratio, AR. (a) 
	Finite aspect ratio correction for the dimensionless flow speed of the trapezoidal sheets studied in Section~\ref{trapezoid}. Comparison of (\ref{Prandtl}) to numerical results obtained with VLM. The 2D limit ($\mathrm{AR}, H/L\rightarrow\infty$) is also shown.}
	\textcolor{black}{(b) Theory for the divergence instability of a trapezoidal inverted flag of infinite AR.} Dimensionless flow speed, $\kappa_\mathrm{lower}^{\infty}$, as a function of the shape parameter, $\Delta/H$, for an inverted flag of trapezoidal morphology (no petiole). \textcolor{black}{Equation (\ref{trapinfty}) holds for $\Delta/H \lesssim O(10)$, and the asymptotic formula in (\ref{traplarge}) is derived for $\Delta/H \gg 1$.} This theory is valid for $H/L \rightarrow \infty$, while measurements are performed on sheets of finite $H/L\sim O(1)$.}
	\label{trapinftyfig}
\end{figure}

Figure~\ref{trapinftyfig}b provides numerical results for $\kappa_\mathrm{lower}^{\infty}$ as a function of the trapezoidal shape parameter, $\Delta/H$, showing that the flow speed at which divergence occurs increases with decreasing $\Delta/H$.
For $\Delta/H=0$, corresponding to a triangular sheet whose  leading edge is a point, (\ref{trapinfty}) gives $\kappa_\mathrm{lower}^{\infty}=6.06$. In the opposite limit, $\Delta/H\rightarrow\infty$, which is an inverted flag with a very wide leading edge relative to its clamped end---i.e., a reversed triangle---(\ref{trapinfty}) predicts $\kappa_\mathrm{lower}^{\infty}=0.549$.
\textcolor{black}{However, the mode shape in (\ref{cantdef}) obviously does not hold when $\Delta/H\gg1$.
In this case, the inverted flag will deflect in a linear (and rigid body) fashion with a small boundary layer near the clamp that provides the flag's elasticity. Any flow speed will deflect the flag in this limit giving $\kappa_\mathrm{lower}^{\infty} \rightarrow 0$ as of  $\Delta/H\rightarrow \infty$, contrary to the above result for a reversed triangle of $\kappa_\mathrm{lower}^{\infty}=0.549$. This case can also be handled using the present theory by replacing the mode shape in (\ref{cantdef}) by that for a trapezoidal plate; which may be approximated using Euler-Bernoulli beam theory~\citep{Landau1970} with a point load at its leading edge. Using this mode shape to evaluate the numerator of (\ref{kappavariational}) and implementing a linearly varying mode shape (rigid body rotation pivoted at the clamp) for its denominator---because it requires solution for the hydrodynamic pressure---gives the following expression,
\begin{equation}
\kappa_\mathrm{lower}^{\infty}=\frac{24 \left(\frac{\Delta}{H} -1\right)^3}{\pi \left(4 +5   \frac{\Delta}{H}\right) \left[2 \left(\frac{\Delta}{H}\right)^2 \log \left(\frac{\Delta}{H} \right)+ \frac{\Delta}{H}\left(4-3
   \frac{\Delta}{H} \right) -1\right]}, \qquad   \frac{\Delta}{H} \gg 1.
\label{traplarge}
\end{equation}
This formula is also plotted in figure~\ref{trapinftyfig}b and shows a natural continuation to very large $\Delta/H$ of the expression in (\ref{trapinfty}), which is valid only up to moderately large  $\Delta/H$. Equation~(\ref{traplarge}) gives the following result in the asymptotic limit, $\Delta/H \rightarrow\infty$,
\begin{equation}
\kappa_\mathrm{lower}^{\infty}\sim\frac{24}{5 \pi  \left[2 \log \left(\frac{\Delta}{H}\right)-3\right]}.
\label{traplargeasym}
\end{equation}
}

\textcolor{black}{We can perform a similar analysis for  $\Delta/H=0$, corresponding to a triangular sheet whose  leading edge is a point. Applying a point load to this leading edge gives $w(x,z)=(1-x)^2$, according to Euler-Bernoulli beam theory. Calculating the hydrodynamic pressure using (\ref{kornecki}) and substituting the result into (\ref{kappavariational}), yields $\kappa_\mathrm{lower}^{\infty}=64/(3 \pi) = 6.79$. This is larger than the original result in (\ref{trapinfty}) of $\kappa_\mathrm{lower}^{\infty}=6.06$ that makes use of the mode shape in (\ref{cantdef}). Since the Rayleigh quotient in  (\ref{cantdef}) guarantees an upper bound~\citep{Leissa1969}, this establishes that $\kappa_\mathrm{lower}^{\infty}=6.06$ as plotted in figure~\ref{trapinftyfig}b is more accurate.
}

\textcolor{black}{Finally, we note that the effects of finite aspect ratio for a trapezoidal sheet can be included through use of  (\ref{Prandtl}), by noting that (see (\ref{AR})):
\begin{equation}
\mathrm{AR} \equiv 2 \frac{\mathrm{max}\left(\Delta,H\right)^2}{\left( H+\Delta \right) L},
\label{ARtrap}
\end{equation}
which increases numerical values for the flow speed at divergence obtained using the formulas in this section.}

\textcolor{black}{The above-described general trend of increasing $\kappa_\mathrm{lower}^{\infty}$ with decreasing $\Delta/H$ agrees with the measurements of $\kappa_\mathrm{lower}$ in figure~\ref{lamina_results}a, which is explored in detail in the next section.}

\subsection{Comparison of theory and measurement}\label{measuredetail}

Figure~\ref{theory} provides a comparison of the measurement data of Section~\ref{trapezoid} for the divergence instability (at which large-amplitude flapping begins) to the theory of Section~\ref{trapezoidformula} \textcolor{black}{in (\ref{Prandtl}), (\ref{trapinfty}) and (\ref{ARtrap})}. Data for two different flag thicknesses, and hence added mass parameters $ \mu = 0.40$ and $0.61$, are reported. This allows assessment of the expected independence of the divergence instability on $\mu$. The divergence instability is chosen as the flow speed at which the flapping amplitude is half of its maximum value, while error bars indicate flow speeds at which the  flapping amplitude is 25\% and 75\% of its maximum value. These error bars specify the uncertainty in quantifying $\kappa_\mathrm{lower}$ for a single inverted flag, and do not include errors arising from the experimental setup, e.g., due to finite width of the clamp, small variations in intrinsic curvature and angle-of-attack across different flags. These additional errors are likely the reason why observations for the two different mass ratios differ slightly. \textcolor{black}{The cusp at $\Delta/H=1 $ is real~\citep{Jones1990,anderson06b} and is also seen in VLM simulations; see figure~\ref{trapinftyfig}a.}

Our theory for $\kappa_\mathrm{lower}$ exhibits the same qualitative behaviour as measurements and increases in a nonlinear fashion for decreasing shape parameter, $\Delta/H$. The 2D strip model of (\ref{trapinfty}) strongly underestimates the measured data, which when corrected using (\ref{Prandtl}) (for finite AR) yields good agreement. Notably, this finite aspect ratio theory slightly overestimates the measurements, which could be due to two independent effects. First, the strip theory of Section~\ref{strip} uses an approximate mode shape and thus strictly provides an upper bound on $\kappa_\mathrm{lower}^{\infty}$, as discussed. This effect is expected to be small given the good agreement with the exact result for a rectangular planform. Second, the Prandtl lifting line correction in (\ref{Prandtl}) implicitly assumes the vortex lines are parallel to the sheet. This is also an approximation and ignores the nonlinear lift generated by the true non-parallel nature of the vortex lines which increases lift~\citep{bollay1939a,sader16a0}. The omission of this (nonlinear lift) factor leads to (\ref{Prandtl}) overestimating the true flow speed at which divergence occurs, as was demonstrated for an inverted flag of rectangular planform by \cite{sader16a0}. The present measurements for a trapezoidal planform are entirely consistent with ~\cite{sader16a0}, showing that the theoretical model for $\kappa_\mathrm{lower}$ in Sections~\ref{strip} and \ref{theoryAR} robustly predicts the divergence instability initiating large-amplitude flapping as a function of inverted flag morphology.


\section{Rectangular inverted flag with a petiole}
\label{rectangle}

Finally, we examine the effect of adding a \textcolor{black}{slender petiole to an inverted flag with a much wider} rectangular planform. The dimensions and properties of these flags are detailed in Table~\ref{table}, and a total of 5 different inverted flags are studied; Flags A -- E; unlike the other flags, Flag E contains no petiole and is measured for reference only. Measurement of each petiole alone, i.e., without its elastic sheet attached, in the wind tunnel shows that the flow has no perceivable effect on its motion at any obtainable flow speed. This establishes that the petiole merely acts as an elastic support---in essence, an end-loaded cantilever---that only exhibits a bending mode. This property is used in the discussion and theory that follows.
\begin{figure}
	\centering
	\centerline{\includegraphics[width=1\columnwidth]{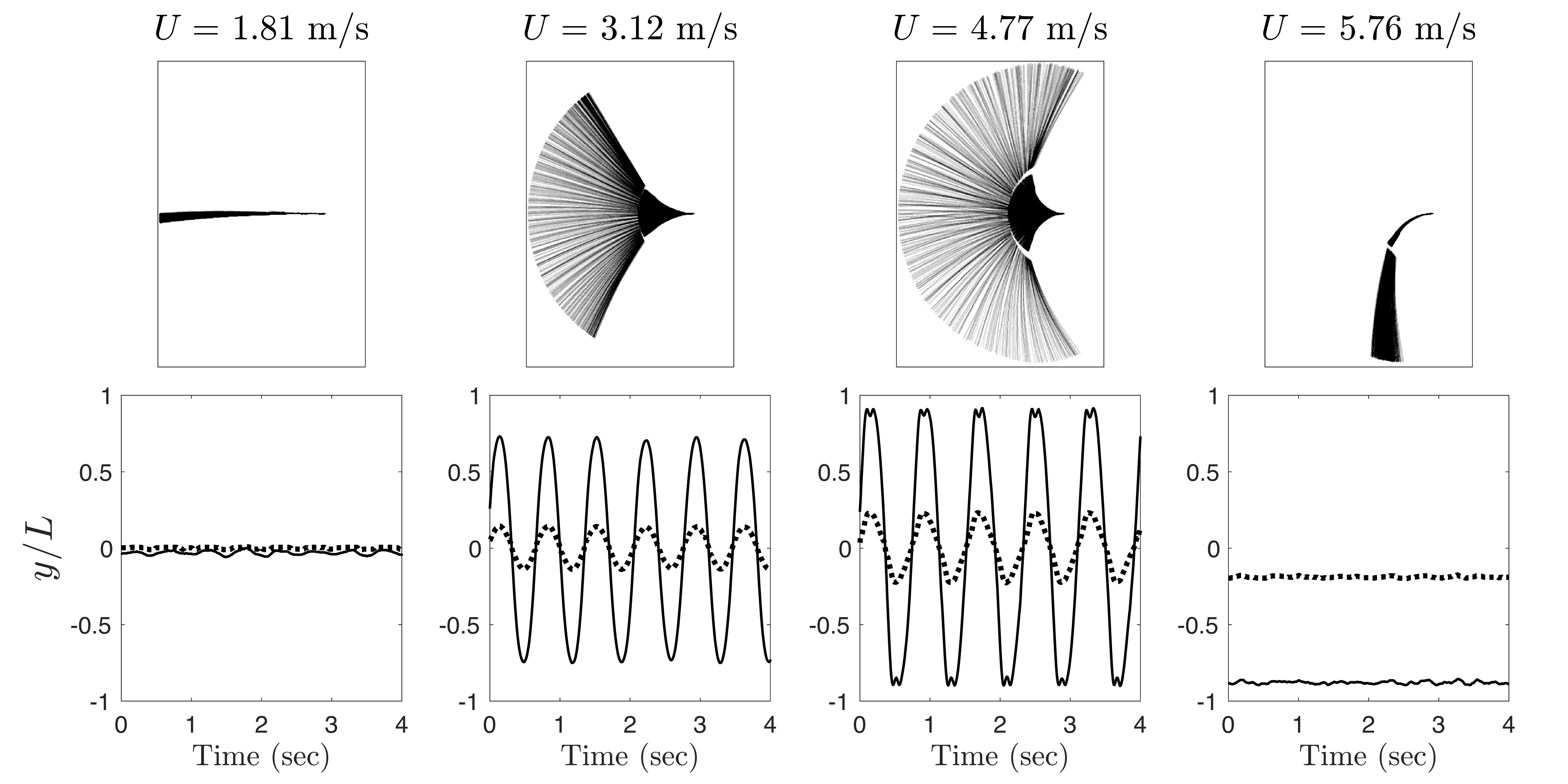}}
	\caption{Stroboscopic images (upper panels) of Flag B at various flow speeds and corresponding time series (lower panels) of the $y$-coordinate of the elastic sheet's free end (solid lines) and petiole-sheet junction point (dotted lines).}
	\label{prog}
\end{figure}

\subsection{General observations}

A sample of measurements on Flag B is given in figure~\ref{prog}, which shows stroboscopic images and time series of this flag's motion as a function of the impinging flow speed, $U$. All major behavioural regimes observed previously for  inverted flags without a petiole are still found. Namely as the flow speed increases, the flag transitions from a straight (undeflected) mode to a flapping mode (periodic and chaotic), and finally, a deflected mode.

All flags exhibit deformation primarily in the petiole, i.e., the elastic sheet moves predominantly as a rigid (and straight) body. This is in contrast to flags without a petiole, where deformation occurs throughout the entire elastic sheet, \textit{cf.} figures~\ref{flap}c and \ref{prog}. These observations establish that the petiole can provide the primary source of elasticity to the flag when present, behaving as an elastic hinge.
Reducing the petiole length, $L_p$, or increasing the petiole rigidity beyond a critical value results in mechanical failure of the elastic sheet, where large stresses at its junction point with the petiole eventually tear the polycarbonate sheet over a few flapping cycles. This observation suggests that damage from flapping in an inverted orientation could impose structural constraints in real leaves. Indeed, it may be a reason why the petiole often extends into the leaf's lamina and gradually tapers off (rather than abruptly ending), while veins into the lamina provide further structural support. 

\begin{figure}
	\centering
	\centerline{\includegraphics[width=1\columnwidth]{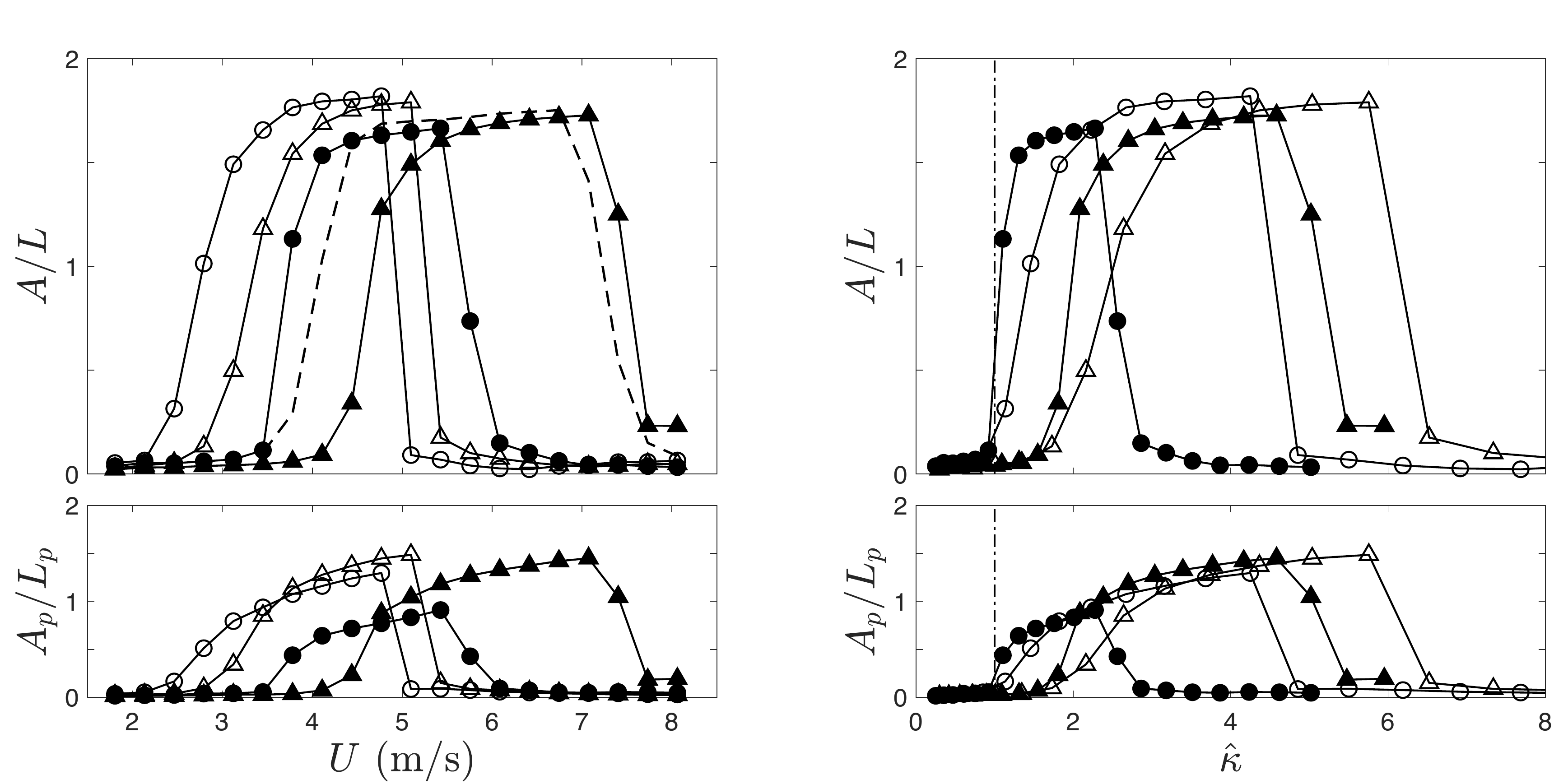}}
	\caption{Dimensionless (peak-to-peak) flapping amplitude of the elastic sheet's free end (top panels) and at the petiole-sheet junction point (bottom panels) for Flag A ($\bullet$), Flag B ($\circ$), Flag C ($\blacktriangle$), Flag D ($\vartriangle$), and Flag E (dashed line). Left panels: Raw data plotted as a function of  the dimensional flow velocity, $U$. Right panels: Same data as in left panels plotted versus the rescaled dimensionless flow speed, $\hat{\kappa}$, defined in (\ref{rescaledpetiole}). This rescaling is theoretically predicted to collapse data for the initiation of large-amplitude flapping onto $\hat{\kappa} = 1$, which is the divergence flow speed in (\ref{flatplatekappa}); shown as vertical dash-dotted line. \textcolor{black}{Experimentally measured values for the dimensionless flow speeds: $\kappa_\mathrm{lower} = 0.038 \pm 0.007$, $0.050 \pm 0.012$, $0.106 \pm 0.015$ and $0.124 \pm 0.037$ for Flags A, B, C and D, respectively. Error bars  quantify the uncertainty in measuring $\kappa_\mathrm{lower}$, and define the 25\% and 75\% points discussed in Section~\ref{measuredetail}.}}
	\label{aonl}
\end{figure}

Figure~\ref{aonl}a gives measurements of (i) the (peak-to-peak) amplitude, $A$, at free end of the elastic sheet, and (ii) the (peak-to-peak) amplitude, $A_p$, at the end of the petiole, for each flag as a function of the impinging flow speed, $U$. The dynamics of the amplitudes, $A$ and $A_p$, align with respect to the flow speed, $U$. This is consistent with the flag deforming in a monotonically increasing fashion from its clamp to its free end, as seen in figure~\ref{prog}. 
Flag C, with the longest and most rigid petiole relative to its elastic sheet, requires the highest critical flow speed to initiate and cease flapping. This flag also displays the highest natural resonant frequency and flapping frequency. Interestingly, large-amplitude flapping in Flag E (that does not contain a petiole) is initiated and stops at comparable flow speeds to those of Flag C. At first sight, this appears to coincide with the flapping and natural resonant frequencies of these two flags being similar; see Table~\ref{table}. However, Flag A also contains similar frequencies but displays very different behaviour; see figure~\ref{aonl}. This demonstrates that the inverted flag's stability is not determined solely by the presence of a petiole nor by its flapping and natural resonant  frequencies, with a more detailed analysis being required.

\subsection{Mechanism for the initiation of flapping}

The above observations that the petiole  (i) is unaffected by the impinging flow, and (ii) provides the primary source of elasticity to the flag, suggests a decoupling of the hydrodynamic and elastic properties of these flags.
This in turn enables the development of a simple theoretical model for the divergence instability that makes use of the lift generated by a rigid flat sheet at small angle-of-attack.  It remains to be seen if such a divergence instability initiates large-amplitude flapping, as found for flags without a petiole. The  theory developed in this section enables this issue to be explored.

The dimensional pressure jump across a flat sheet of length, $(L-L_p)$, at angle-of-attack, $\alpha$, to an impinging flow of speed, $U$, \textcolor{black}{follows from (\ref{kornecki}) and is given by}
\begin{equation}
p=2 \rho U^2 \alpha \sqrt{\frac{L-L_p}{x}-1},
\label{pressureflat}
\end{equation}
where $x$ is the dimensional coordinate in the flow direction, whose origin is at the leading edge of the rigid sheet; see figure~\ref{petiole}a. This rigid sheet is attached to the petiole, a cantilevered elastic beam of length $L_p$, whose deflection function, $w(x)$, obeys the Euler-Bernoulli equation~\citep{Landau1970}. It then follows that the force, $F$, and moment, $M$, exerted by the flat plate on the end of the petiole are
\begin{equation}
\label{loadsflatplate}
F= \pi \rho U^2 \! H \! \left(L-L_p\right) w'(L_p),  \qquad
M= -\frac{3\pi}{4}\rho U^2 \! H\! \left(L-L_p\right)^2 w'(L_p) ,
\end{equation}
where the rigid sheet's angle-of-attack, $\alpha \equiv - w'(L_p)$. \textcolor{black}{The small tab-like overlap hinge connecting the sheet and petiole has its own bending stiffness, which is ignored in the present model, i.e., this hinge is assumed to be infinitely stiff.}

The Euler-Bernoulli equation for this end-loaded beam is $w''''(x)=0$, with boundary conditions, $w(L)=w'(L)=0$, $w''(L-L_p)=-M/EI$ and $w'''(L-L_p)=F/EI$, where $EI = E h_p^3 H_p/12$. Solving this eigenvalue problem gives the required dimensionless flow speed for which divergence occurs:
\begin{equation}
\kappa_\mathrm{lower}^\mathrm{pet}\equiv\frac{\rho U_\mathrm{lower}^2 L_p^2 \left(L-L_p\right)}{D'}\approx\frac{4}{\pi} \left(1+\frac{2}{\mathrm{AR}}\right)  \frac{L_p}{3 L - L_p} \frac{H_p}{H}  ,
\label{flatplatekappa}
\end{equation}
where $D'\equiv E h_p^3/12$ is the flexural rigidity of the petiole, $U_\mathrm{lower}$ is the critical flow speed for divergence, and the Prandtl's lifting line correction, (\ref{Prandtl}), for the sheet's finite generalised aspect ratio, AR, is included, with
\begin{equation}
\mathrm{AR}=\frac{H}{L-L_p} .
\label{extra}
\end{equation}
We emphasise that (\ref{flatplatekappa}) assumes that the flag's elasticity is isolated to the petiole; see above. 

Figure~\ref{aonl}b presents the same data as in figure~\ref{aonl}a but now plotted as a function of the rescaled dimensionless flow speed,
\begin{equation}
\hat{\kappa} \equiv   \frac{\kappa'}{\kappa_\mathrm{lower}^\mathrm{pet}} ,
\label{rescaledpetiole}
\end{equation}
where $\kappa'\equiv\rho U^2 L_p^2\left(L-L_p\right) /D'$. Equation~(\ref{rescaledpetiole}) implicitly contains the divergence instability's geometric dependence in (\ref{flatplatekappa}) for an inverted flag with a petiole; divergence is theoretically predicted to occur at $\hat{\kappa}\approx1$. Data for Flag E is omitted from figure~\ref{aonl}b because it does not have a petiole.

\begin{figure}
	\centering
	\centerline{\includegraphics[width=0.6\columnwidth]{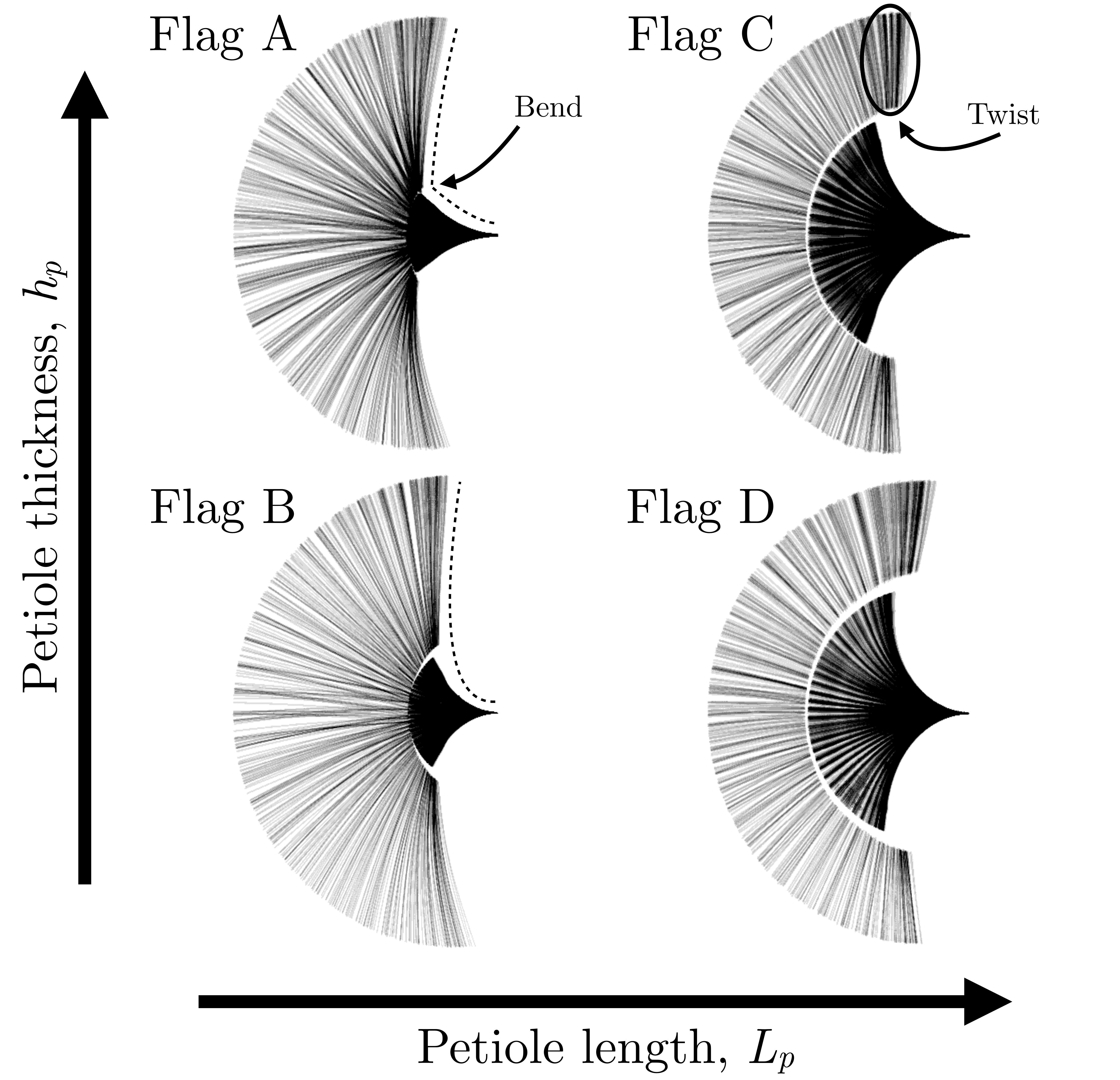}}
	\caption{Stroboscopic images of Flags A -- D in their respective flapping regimes at various flow speeds, which are selected so that the flags exhibit similar flapping amplitudes. Flag A: $U=4.77$ m/s. Flag B: $U=4.11$ m/s. Flag C: $U=6.09$ m/s. Flag D: $U=4.77$ m/s. Flag A exhibits a notable kink where the petiole connects the elastic sheet, which is smaller than that for Flag B (as indicated by dashed lines). Flags C and D display significant twisting in their elastic sheets, which is manifest by an apparent discontinuity at the connection between the sheet and petiole---the images are taken in the gravity direction.}
	\label{flagsABCD}
\end{figure}

This rescaled plot shows that (\ref{flatplatekappa}) accurately captures the divergence instability for Flag A---with data collapsing onto $\hat{\kappa}\approx1$ in the flow speed region where large-amplitude flapping is initiated (at low flow speed). This rescaling also works well for Flag B, although divergence (and the initiation of flapping) occurs at a slightly higher value of $\hat{\kappa}$. The reason for this difference is evident in stroboscopic images of these flags in the flapping regime; see figure~\ref{flagsABCD}. While the theory derived in this section assumes the deflection angle of the petiole and elastic sheet are identical at their connection position, this is not satisfied exactly in measurements: Flags A and B displaying different behaviour at this connection point. The thicker (and hence more rigid) the petiole the stronger the discontinuity, which leads to a larger violation of the theory's assumptions for Flag B relative to Flag A; \textcolor{black}{the hinge connecting the sheet and petiole can deform to a greater degree, violating the model's assumption of an infinitely stiff hinge.} Even so, this is found to exert a small effect with the derived approximate theory performing well and accurately capturing the divergence instability for both Flags A and B with no fitting parameters. \textcolor{black}{We emphasise that the theory contains a number of approximations---in addition to the rigid hinge assumption---and therefore precise agreement between measurement and theory is not expected.}

In contrast, the divergence instability and onset of flapping is underestimated by the developed theory for Flags C and D, with divergence occurring at $\hat{\kappa}\approx2$ ($>$1). The stroboscopic images in figure~\ref{flagsABCD} also provide some insight into this difference, with strong twisting behaviour in Flags C and D. This is manifest in an apparent discontinuity in the displacement between the petiole and the sheet, which is greatly reduced in Flags A and B. \textcolor{black}{Such behaviour, due to the coupling effects of gravity and twisting of the petiole (the sheet deforms in a rigid fashion with the inverted flag displaying a waving motion as it flaps),} is not included in (\ref{flatplatekappa}) and appears to be causing the divergence instability to occur at twice the theoretical value. Even so, increasing the petiole thickness enhances the value of $\hat{\kappa}$ at which divergence is observed, as for Flags A and B.
\textcolor{black}{Similar twisting behaviour has been previously observed for slender inverted flags~\citep{sader16a0}. Reducing the petiole length enhances its torsional stiffness and decreases this behaviour; see figure~\ref{flagsABCD}.}

For Flags A -- D, the undeflected mode of the inverted flag abruptly gives way to large-amplitude flapping. This shows that a divergence instability triggers this flapping instability, as observed for inverted flags without a petiole.

Finally, we note that the rescaling in (\ref{rescaledpetiole}) does not collapse data at higher flow speed where flapping ceases and the deflected mode appears. Thus,  the empirical relation of $\kappa_\mathrm{upper}/\kappa_\mathrm{lower}\approx 4$  observed for trapezoidal inverted flags with no petiole (Section~\ref{trapezoid}) does not hold for rectangular flags with petioles. Adding a petiole to an inverted flag can therefore enable adjustment of the relative flow speed range over which large-amplitude flapping occurs.


\section{Conclusions}

Motivated by the form and occurrence of natural leaves, we have studied the effect of morphology on the stability and dynamics of inverted flags. This involved a combination of experiment and theory with good agreement observed throughout.
Two approaches were taken to examine this effect: (i) changing the shape of the inverted flag's elastic sheet using a trapezoidal morphology, and (ii) adding a petiole to the widely studied rectangular elastic sheet. This showed that the inverted flag flapping instability, as first reported by \cite{kim13a}, is highly robust and persists across a wide range of morphologies.
 
Trapezoidal inverted flags (without a petiole) were found to exhibit large-amplitude flapping that depends strongly on their morphology, $\Delta/H$. Strikingly, it was observed that the ratio of the flow speed where flapping ceases to that of its onset is $\kappa_\mathrm{upper}/\kappa_\mathrm{lower}\approx 4$ over the entire geometric range  studied, $0 \le \Delta/H \le 1.75$; it remains to be seen if a similar result holds for other morphologies.
A general theoretical model for the onset of a divergence instability, which initiates flapping at low flow speeds, was presented in Section~\ref{theorysec} for inverted flags of arbitrary morphology. This was applied to the above-mentioned trapezoidal inverted flags, yielding an analytical solution that was   found to exhibit good agreement with measurement.

Second, inclusion of a petiole to inverted flags with rectangular morphologies was observed to result in rigid body motion of the flag's elastic sheet, with the petiole providing the requisite elasticity---thus, decoupling hydrodynamic and elastic forces. This motivated development of a simple analytical model for the divergence instability of these flags. Twisting of the sheet was observed for some flags, and linked to an increase in the flow speed at which divergence occurred. Further work may examine the effect of coupling the elastic sheet's morphology to that of a petiole and non-idealities in the dynamics of the flag's motion.

Returning to a biological setting, the white poplar leaf combines a roughly triangular lamina with a long petiole, two features that our study suggests have counteracting effects. The former would increase the critical flow speed for the onset of flapping while the latter would decrease it. Our work therefore motivates wind tunnel experiments on a variety of real tree leaves that could potentially shed light on whether there exists an optimal leaf morphology to resist wind damage in the inverted orientation. Finally, a leaf is not an isolated system in nature but rather part of a larger system of interacting leaves and branches. This study therefore represents only but a first step in understanding such a complex system, which may exhibit new and unexplored phenomena.


\section*{Acknowledgments}

We acknowledge funding from the Gordon and Betty Moore Foundation, the Lester Lees Aeronautics Summer Undergraduate Research Fellowship, the Australian Research Council Centre of Excellence in Exciton Science (CE170100026) and the Australian Research Council grants scheme.

\bibliographystyle{jfm}

\end{document}